\documentclass[11pt,aps,prd,preprint,preprintnumbers,showpacs,superscriptaddress,floatfix]{revtex4-1}
\usepackage{graphicx}
\usepackage{epsfig}
\usepackage{dcolumn}
\usepackage{bm}
\usepackage{subfigure}
\usepackage{amsmath}
\usepackage{amssymb}
\usepackage{multirow}
\usepackage{setspace}
\usepackage{xcolor}
\usepackage[colorlinks,
            linkbordercolor={0 0 0},
            pdfborder={0 0 0},
            linkcolor=blue,
            citecolor=blue,
            urlcolor=blue,
            breaklinks=true]{hyperref}

\allowdisplaybreaks

\newlength{\x}
\newlength{\y}
\newlength{\z}
\phantomsection

\begin{document}

\title{ Surface effects on hydrodynamic evolution}

\author{Sanatan Digal}
\email{digal@imsc.res.in}%
\affiliation{The Institute of Mathematical Sciences, Chennai, 600113, India}%
\affiliation{Homi Bhabha National Institute, Training School Complex,
Anushakti Nagar, Mumbai 400085, India}%

\author{P.S. Saumia}%
\email{saumia@theor.jinr.ru}%
\affiliation{Bogoliubov Laboratory of Theoretical Physics, JINR, 141980 Dubna, \
Russia}%
\begin{abstract}
We study the effect of surface tension of the phase boundary in the dynamics of an expanding fluid.
A fluid at local thermal equilibrium, but has a slowly varying temperature profile, like
the plasma formed in heavy ion collisions, will have rapidly varying order parameter 
field at the edge
of the plasma where the temperature falls below the transition temperature. In the case where the
free energy admits a first order transition, the gradient
energy of this field will act as surface tension. We couple
hydrodynamics and order parameter field evolutions to study the effect of this surface
in the expansion of the plasma. We see that the surface slows down the expansion
which reflects in the development of radial flow and momentum anisotropy.  
	
\end{abstract}
\maketitle


\section{Introduction}
\label{sec:intro}  

Heavy ion collision experiments provide us with a way to study the properties 
of strongly interacting matter at extreme conditions. One of the important observations
from high energy heavy ion collisions is the collective flow. It indicates that the
matter formed at these collisions reach thermodynamic equilibrium rapidly and evolves as a fluid, 
expand and cool down. Subsequently the constituent particles freeze out before reaching the detectors. 
The various 
stages of the evolution can be indirectly observed from the flow of different 
particles that freeze out at different times. The high energy collisions at RHIC and LHC 
creates matter in the deconfined regime at high temperatures and very small baryon densities.
Lattice studies suggest that the transition from deconfinement to confinement of 
strongly interacting matter at these conditions is a cross over~\cite{crsovr}.  Various
theoretical models predict a critical end point followed by a first order transition line
 at higher baryon densities~\cite{crfo}. There are attempts 
to study higher baryon density regimes at experiments like FAIR and NICA and one expects
to see some indications of a first order phase transition at these experiments~\cite{nica, fair}.  

There have been several approaches to understand the dynamics of the quark gluon plasma 
formed in heavy ion collisions. Hydrodynamics with an appropriate
equation of state has been successful in explaining the flow observables at RHIC and LHC~\cite{flowrev,flowrev2}.
There also are attempts to understand the dynamics of the phase transitions
as well as the possibility of formation of center domains and topological defects 
using the evolution of order parameter fields using different effective field theory models~\cite{rkmams1,rkmams2,rkmamsfo,Mishra:2016ipq}. 
First order phase transition dynamics has also been studied by reducing the hydrodynamics 
equations into equations for order parameter of the system in the vicinity of the 
critical point~\cite{dnv}.
But even when the system is well in the deconfined region, in heavy ion collisions,
the system will have a spatial profile of temperature slowly varying away from the centre to 
zero towards the edges. Hence the system at all times after thermal equilibration should have a 
phase boundary until it cools down below the transition temperature. This means that the order 
parameter at this boundary will rapidly vary from deconfinement to confinement value and the 
order parameter field will have a gradient energy.
In the case of a first order equation of state, this gradient energy will always be finite
and will act as surface tension. The radial expansion of the plasma in this case should feel 
a resistance since increasing the surface area of the system increases the surface energy. 
Intuitively, then it suggests that the system with a phase boundary should have a constrained 
radial expansion compared to a system without a phase boundary.  
 
Hydrodynamics alone, even with a first order equation of state, cannot address surface
tension effects at the boundary. Effective field theory with an appropriate order 
parameter, on the other hand, can be used to study the evolution of the phase boundary where the effect 
of surface tension is automatically taken care of, but cannot be used to study collective flow. 
Hence, in this work, we attempt to couple these
two together, a fluid in the bulk with a surface which is represented by the gradient energy of an 
order parameter field. The evolution of the coupled system is expected to handle the 
surface effects on the flow of the fluid.
We would like to clarify that we do not study the dynamics of the phase transition of the system through 
bubble nucleation here. Our system is in the deconfined region but it has a boundary across which the temperature 
falls below the transition temperature and hence has a surface.  

In section \ref{sec:form}, we discuss the formulation of the problem and explain the 
Polyakov loop model which is used for the evolution of order parameter here. In section \ref{sec:imp},
the equation of state and initial conditions are explained along with the implementation of the
techniques used to numerically solve the partial differential equations. Section \ref{sec:results}
discusses the results which show the effects of phase boundary on the development of radial
flow as well as momentum anisotropy of the fluid and section \ref{sec:summ} summarizes the study.

\smallskip

\section{Coupling the Polyakov loop field to the fluid}  
\label{sec:form} 

To begin with, let us assume that we have two systems interacting with each other, one is
the fluid and the other is an order parameter field. Then the evolution of
the system is governed by the conservation of the energy momentum tensor of the
combined system. This should be supplemented by an equation of state for the fluid as well
as an equation of motion for the evolution of the order parameter field.
 For confinement deconfinement transition, Polyakov loop $\phi$ is a 
good order parameter. For simplicity, we assume that the fluid is quark less. Then we can use
a Polyakov loop effective field theory model for the equation of state as well as for the
dynamics of the order parameter part the system. We use the model by Pisarski \cite{Pisarski:2000eq,
 Dumitru:2000in, Dumitru:2001vc} which
has a potential with a weak first order transition. Thus the
equations we need to solve are:
\begin{equation}
\partial_{\mu}T^{\mu \nu} = 0
\label{consrv}
\end{equation}
and
\begin{equation}
\frac{\partial^2 \phi}{\partial t^2} - \nabla^2 \phi = -\frac{\partial V(\phi,T)}{\partial \phi}
\label{EulLag}
\end{equation}
where  
\begin{equation}
T^{\mu \nu} = T^{\mu \nu}_{fluid} + T^{\mu \nu}_{\phi}
\end{equation}
Eq. (\ref{EulLag}) describes the evolution of the order parameter field given an initial 
condition~\cite{2ndorder}.
$T^{\mu \nu}_{fluid}$ is the energy-momentum tensor of the fluid given by
\begin{equation}
T^{\mu \nu}_{fluid}=(\epsilon + p)u^{\mu}u^{\nu}+p\eta^{\mu \nu}
\end{equation}
with $\epsilon$ and $p$ the energy density and pressure density of the fluid, $u^{\mu}=\gamma(1,\vec{v})$
the 4-velocity and $\eta^{\mu \nu}$ the Minkowski metric with signature (- + + +). $\vec{v}$ is the 
3-velocity of the fluid element and $\gamma$ is the corresponding Lorentz factor given by 
$\gamma = \frac{1}{\sqrt{1-\vec{v}^2}}$.
$T^{\mu \nu}_{\phi}$ is the energy-momentum tensor of the Polyakov loop field $\phi$ and 
$V(\phi,T)$ is the effective potential at temperature $T$ given 
by
\begin{equation}
	V(\phi,T)  = b_4 T^4 \left[ -{{b_2 (T)} \over 4} 
	\phi^2 - {{b_3 } \over 6} \phi^3  
+ {{1} \over 16} \phi^4 \right].
\label{polpot}
\end{equation}
Note that the Polyakov loop field is usually written as a complex scalar field 
and the potential should be written as a function of $\phi$ and $\bar{\phi}$. But
here we ignore fluctuations of the order parameter field and assume that the 
imaginary component of the field is not excited or remains small during the evolution. 
Thus we can write $\bar{\phi} = \phi$
and that simplifies our implementation. The above 
effective potential Eq.(\ref{polpot}) with the following form of $b_2 (T)$
\begin{equation}
{b_2 (T)} = \left(1- 1.11 ~T_c / T \right) \left(1 + 0.265~ T_c / T \right)^2 
 \left(1 + 0.3 ~T_c / T \right)^3 -0.487
\end{equation}
and the coefficients $b_3 = 2.0$ and $b_4 = 0.6016$ reproduces the pressure of the pure gauge 
theory computed from non-perturbative lattice method(s) and is given by $p=-V(\phi,T)$. $T_c$
is the transition temperature.  

Since the order parameter field is a scalar field, the corresponding energy-momentum
tensor can be calculated as follows. The Lagrangian of the field is given by 
\begin{equation}
L(\phi,T) = \alpha  T^2 \partial_{\mu} \phi \partial^{\mu} \phi - V(\phi,T)
\end{equation}
The corresponding energy-momentum tensor is given by
\begin{equation}
T^{\mu \nu}_{\phi} = \frac{\delta L}{\delta(\partial_{\mu} \phi)} - \eta^{\mu \nu} L
\label{tmn}
\end{equation}
After doing the variation of the Lagrangian, $T^{\mu \nu}$ can be written as
\begin{equation}
T^{\mu \nu}_{\phi} = 2 \alpha T^2 \partial^{\mu} \phi \partial^{\nu} \phi - \eta^{\mu \nu} (\alpha T^2 \partial_{\beta} \phi \partial^{\beta} \phi - V(\phi,T))
\label{polT}
\end{equation}
$\alpha$ is a constant given by $2N/g^2$ \cite{Dumitru:2000in}, where $N$ is the number of colors and $g$ is the gauge coupling constant. For $g/4\pi = 0.3$, $\alpha = 1.6$.

Let us look at Eqs. (\ref{consrv}) and (\ref{EulLag}) again. Even though we need to solve these two
equations together, the fluid part with variables like energy density, pressure and
4-velocities seems almost uncoupled to the Polyakov loop field part which only has the field
variable. But there is a mechanism of feedback between these two systems and that is through the 
temperature of the fluid. The Polyakov loop effective potential is a function of temperature and 
hence the field evolution in Eq.(\ref{EulLag}) will depend on the local temperature which is determined 
by the fluid parameters $\epsilon$ and $p$ via the equation of state which in turn can be written 
down from the effective model.

Lattice results are used
to estimate the thermodynamic variables of the Polyakov loop field at a certain temperature 
and these variables  are equated with that of the fluid for hydrodynamic simulations.  Similarly
using Polyakov loop model, at thermodynamic equilibrium, pressure of the fluid is estimated as $-V(\phi,T)$ at the given temperature.  So if the field is 
at the minimum of the potential, adding up the pressure of the fluid and $-V(\phi,T)$ in $T^{\mu\nu}$
 leads to double counting. 
Note that even though fluid pressure can be obtained from the free energy of the Polyakov loop, the 
evolution of the fluid and the Polyakov loop field are very different. For example,
the evolution of the Polyakov loop field hardly shows any flow. This is because, Polyakov loop
captures only the low momentum modes of the system as it is a condensate field. 

If we look at Figs.~\ref{eos} and \ref{polTin}, we can see that the value of the field differs from its value at the minimum
only near the surface. Here the field interpolates between its value at confinement (which is zero)
to its value at deconfinement at the transition. So assuming local thermal equilibrium,
we consider only the fluid pressure and neglect the contribution of the potential $V(\phi,T)$ in
$T^{\mu\nu}_{\phi}$.
Thus the only non-zero term in $T^{\mu\nu}_{\phi}$ is the gradient energy  at the surface and this serves as surface 
tension at the phase boundary. 
Since the field is away from its minimum at the surface due to gradients, $-V(\phi,T)$ will
not be same as the pressure of the fluid here. It is not clear how to calculate the energy 
momentum tensor of the field in this case. One could
write down the pressure contribution from the field as equal to the difference between the 
equilibrium value and the value of the field profile, ie, $-(V(\phi,T)-V_{eq}(\phi,T))$ where 
$V_{eq}(\phi,T)$ corresponds to the minimum value of the free energy at a given temperature. 
This will make sure that it remains zero everywhere
except at the phase boundary. Effectively, this is what we are doing by neglecting the effective
potential contribution in the energy momentum tensor of the Polyakov loop field, except at the surface.  
 But at the boundary even if this difference remains negligible compared to the
gradient term, it gives a negative contribution to the total pressure and it leads to numerical 
errors. Hence in the results presented, this term is ignored in $T^{\mu\nu}_{\phi}$ and only gradient term is considered. 
 
 



\section{Implementation}
\label{sec:imp} 

\subsection{Equation of state}

Using thermodynamic relations, the energy density can be calculated from the free energy as, 

\begin{equation}
\epsilon = -T\left ( \frac{\partial V(\phi,T)}{\partial T}\right )_{volume=constant} + V(\phi,T)
\end{equation}

\begin{figure}[h]
  \centering
  \subfigure[]
  {\rotatebox{0}{\includegraphics[width=0.4\hsize]
      {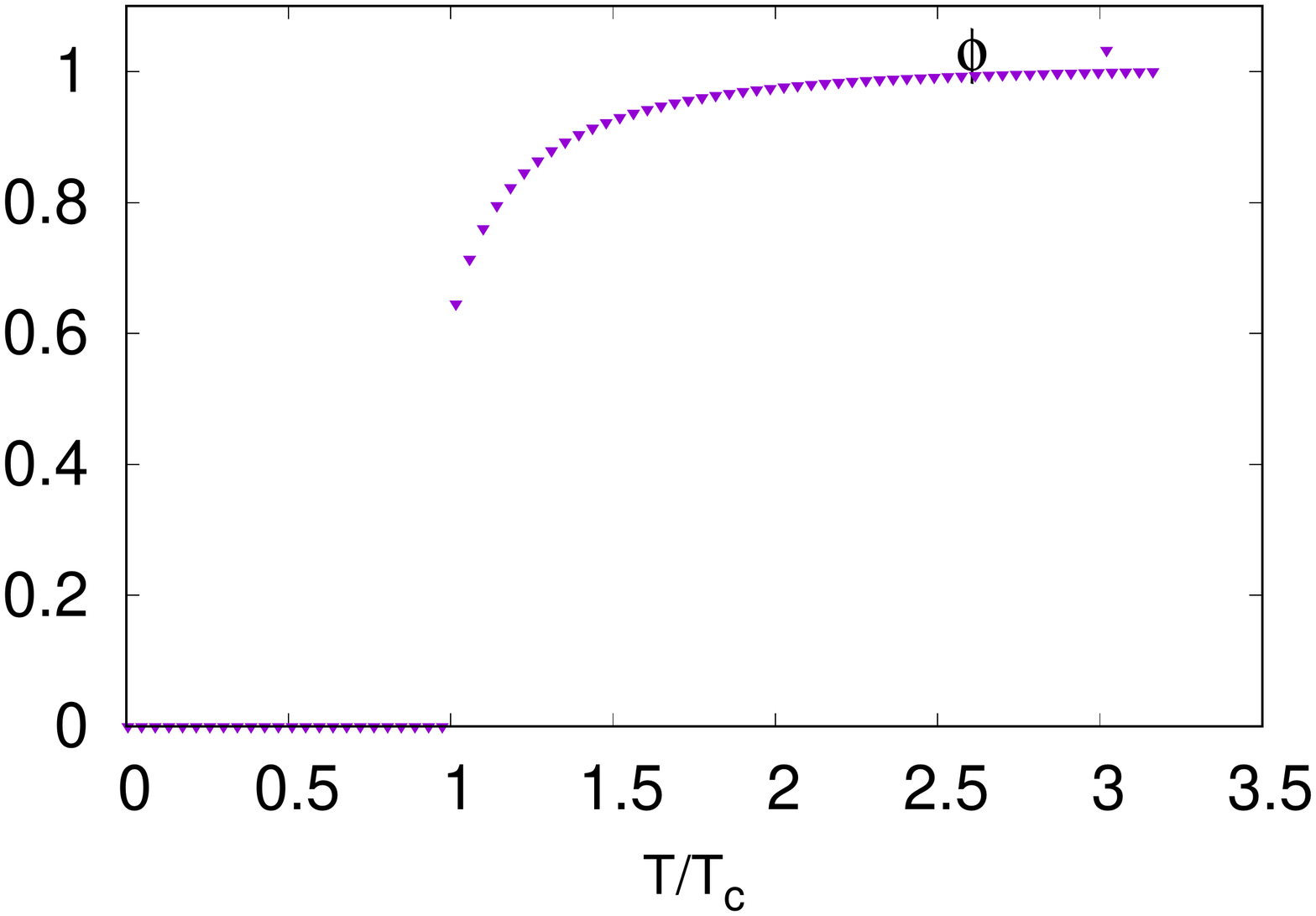}}
      \label{lvst}
  }
  \subfigure[]
  {\rotatebox{0}{\includegraphics[width=0.4\hsize]
      {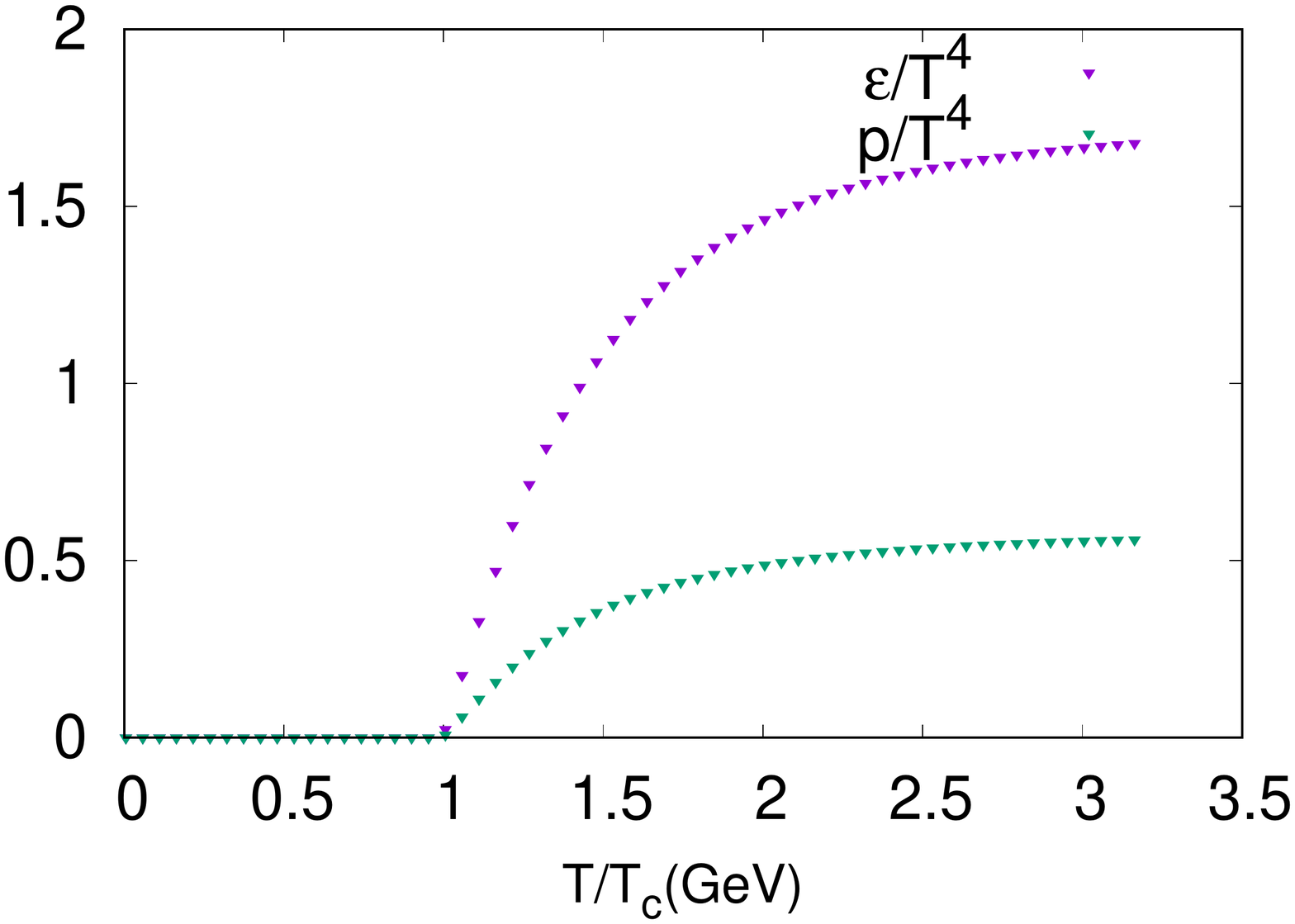}}
      \label{epvst}
  }
\caption{Equation of state}
\label{eos}
\end{figure}
Minimizing the effective potential (\ref{polpot}) with respect to $\phi$ and using the corresponding field value at the 
given temperature, the energy density and pressure can be calculated. This gives the equation of state for the fluid.
Fig. \ref{lvst}
shows the order parameter as a function of temperature and Fig. \ref{epvst} shows the variation of
energy density ($\epsilon/T^4$) and pressure ($p/T^4$). In this model, the order parameter is zero below 
the transition temperature and as a result the pressure remains zero in the confined phase. There are 
several problems because of this when we try to evolve the fluid using this equation of state. These 
problems will be discussed later in detail.

\subsection{Reduction into 2D and initial conditions}
 We don't
 consider Bjorken expansion along the longitudinal axis. Here we have a system with a first order
 transition. First order phase transitions are expected at low temperatures and high baryon densities for which Bjorken 
 flow component will be small. So for central collisions, we have
 a spherical system of expanding fluid and for non-central collisions, we have an ellipsoid. 
 For simplicity we assume that our system has an azimuthal symmetry. 
 Then we can reduce the system into 2+1 dimensions 
 and use cylindrical co-ordinates to solve Eqs. (\ref{consrv}) and (\ref{EulLag}). 
   The equation of motion for the order parameter (\ref{EulLag}) then reduces to

\begin{equation}
\frac{\partial^2 \phi}{\partial t^2} -  \frac{\partial^2 \phi}{\partial r^2} -\frac{1}{r}\frac{\partial \phi}{\partial r} - \frac{\partial^2 \phi}{\partial z^2}= -\frac{\partial V(\phi,T)}{\partial \phi}
\label{EulLag2}
\end{equation}

\begin{figure}[h]
  \centering
  {\rotatebox{0}{\includegraphics[width=0.85\hsize]
      {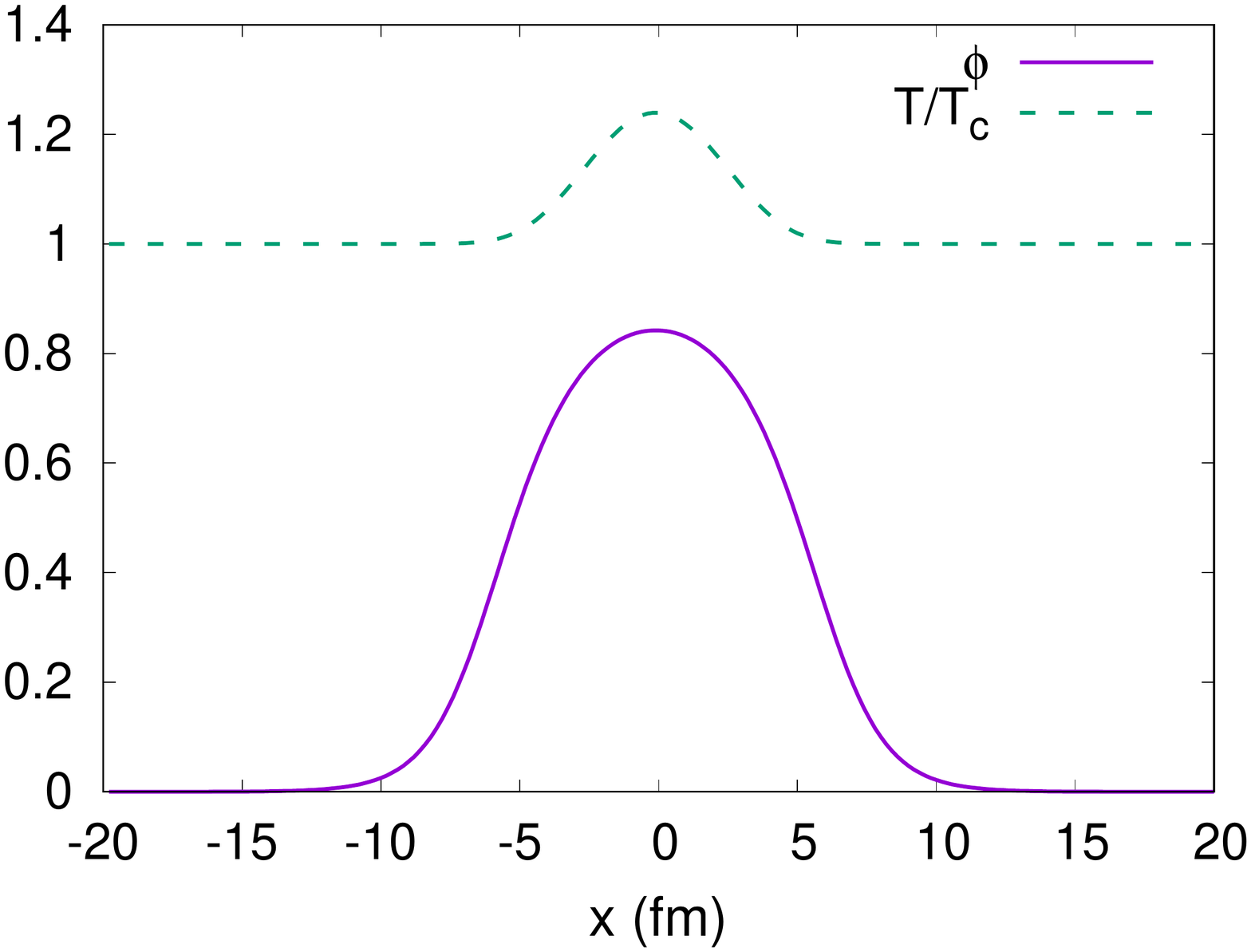}}
  }
\caption{Initial temperature and Polyakov field profile for the coupled system along x-axis.}
    \label{polTin}
\end{figure}

We assume simple Gaussian initial conditions along $r$ and $z$ for the initial energy density of 
the fluid. For non-central collisions, the initial conditions require different widths for the
initial Gaussian along $r$ and $z$. The major axis of the ellipsoid is along $z$ 
in cylindrical coordinate system, so the width along $z$ will be larger.
Using the equation of state, the initial pressure and temperature profiles are obtained. 
The initial $\phi$ field is obtained by evolving Eq.(\ref{EulLag2}) with this initial temperature profile fixed.
We use a small dissipation term for the Polyakov loop field so that it
settles into the minimum of the potential. In the absence of dissipation, the field tends to oscillate 
around the minimum. No matter what initial profile is chosen,  
Eq.(\ref{EulLag2}) along with dissipation evolves it into the one which minimizes the energy with respect 
to the given temperature profile. Fig. \ref{polTin} shows this input Polyakov loop profile for the coupled 
system with the corresponding temperature profile. 

We have seen earlier that energy density and pressure of the Polyakov loop field are zero outside 
the deconfined region for our equation of state. 
This means that the initial Gaussian profile of energy density cannot go to zero at large radii. Any non-zero 
value below the value of the energy density at transition temperature is unphysical. Thus
we let the Gaussian at large radii go to the value of $\epsilon$ at the transition temperature. 
But in order to have a phase boundary, we need a confined phase outside and this is ensured from
the fact that the order parameter is zero in that region. 

\subsection{Solving the PDEs}

The conservation equations \ref{consrv} are rewritten by separating the fluid and the field 
parts as
\begin{equation}
\partial_{\mu}T^{\mu \nu}_{fluid}=-\partial_{\mu}T^{\mu \nu}_{\phi}
\label{consrv2}
\end{equation}
We use Kurganov-Tadmor scheme \cite{kt,Schenke:2010nt} on the fluid part of the conservation equation (\ref{consrv2}) which 
turns it into a set of ordinary differential equations and use a second order Runge-Kutta method to solve it numerically as in 
\cite{Schenke:2010nt}. Here we only quote the final equations of the scheme in one spatial dimension. For more than one 
dimension, the scheme is repeated in each of them. If we have a conservation equation in one spatial dimension and is given by 
\begin{equation}
\partial_t u = -\partial_x f
\end{equation}
where $f$ can be written as $f=vu$, the final result in the $\Delta t \to 0$ limit of this equation can be written as a conservation equation of the cell average 

\begin{equation}
\bar u(t) = \frac{1}{\Delta x} \int_{x_{j-1/2}}^{x_{j+1/2}} dx u(x,t)
\end{equation}
and is given by
\begin{equation}
\frac{d\bar u(t)}{dt} = - \frac{F_{j+1/2}(t)-F_{j-1/2}(t)}{\Delta x}
\label{kt}
\end{equation}
where
\begin{equation}
\begin{split}
 F_{j\pm 1/2} = &\frac{f_{j\pm 1/2,+}(t)+f_{j\pm 1/2,-}(t)}{2} \\
      &-\frac{a_{j\pm 1/2}(t)}{2}\left(\bar u _{j\pm 1/2,+}(t)-\bar u_{j\pm 1/2,-}(t)\right)
\end{split}
\end{equation}
The subscripts $j\pm 1/2$ denote points in space separated by $\Delta x/2$ from the point $j$.
Here $a$ is the maximum propagation speed given by $|\partial f/\partial u|$ and 
 $f_{j\pm 1/2,\pm}$ is calculated from 

\begin{equation}
\bar u_{j+1/2,+}=\bar u_{j+1}-\frac{\Delta x}{2}(u_x)_{j+1}
\end{equation}

\begin{equation}
\bar u_{j-1/2,-}=\bar u_{j}+\frac{\Delta x}{2}(u_x)_{j}
\end{equation}
$(u_x)_j$ is the spatial derivative and is determined by the minmod flux limiter which is given by
\begin{equation}
(u_x)_j=\text{minmod}\left(\theta \frac{\bar u_{j+1}-\bar u_j}{\Delta x},\theta \frac{\bar u_{j+1}-\bar u_{j-1}}{2\Delta x},\theta \frac{\bar u_{j}-\bar u_{j-1}}{\Delta x} \right )
\end{equation}
where 
\begin{equation}
\text{minmod}(x_1,x_2,...) = 
\begin{cases}
\text{min}~x_j, &\text{if} ~ x_j > 0 ~\forall j \\
\text{max}~x_j, &\text{if} ~ x_j < 0 ~\forall j \\
0, ~~~~~~~~~\text{otherwise}
\end{cases}
\end{equation}
$\theta$ is a parameter that controls the numerical diffusion and oscillation and we choose $\theta=1.1$ \cite{Schenke:2010nt}. 

Now Eq.(\ref{kt}) can be solved using an ODE solver. We use Heun's method to solve this 
and the steps for an equation of the form
\begin{equation}
\frac{du}{dt}=F(t)
\end{equation}
are as follows:

\begin{equation}
\text{Heun's}:
\begin{cases}
u^{(1)}=u^n+\Delta t F(t,u^n)\\
u^{n+1}=u^n+\frac{\Delta t}{2}\left( F(t,u^n)+F(t+\Delta t,u^{(1)})\right)
\end{cases}
\end{equation}
The superscripts $n$ and $n+1$ denote consecutive time steps and the index 1 denotes an
intermediate value.
The field contribution on the right hand side of Eq.(\ref{consrv2}) is treated as a source term which is 
handled as in \cite{KTsrc}. There  is a time derivative part in the source term and we follow 
\cite{Schenke:2010rr} and calculate it as $\dot s_t = (s_t-s_{t-\Delta t})/\Delta t$ for the first 
step of Heun's method and use the value of the source term calculated in the intermediate step
 $s^*_{t+\Delta t}$ to determine the time derivative in the second step, ie, $\dot s_t = (s^*_{t+\Delta t}-s_{t})/\Delta t$.

Eq.(\ref{EulLag2}) is rewritten as four first order equations as below (Eqs.(\ref{el1}) to (\ref{el4})) 
and solved using Lax-Friedrichs scheme. We define $\Psi_2=\frac{\partial\phi}{\partial r}$ and
 $\Psi_3=\frac{\partial\phi}{\partial z}$ and

\begin{equation}
\Psi_1 = \frac{\partial \phi}{\partial t}
\label{el1}
\end{equation}

\begin{equation}
\frac{\partial\Psi_1}{\partial t} = \frac{\partial \Psi_2}{\partial r}+\frac{1}{r}\Psi_2+\frac{\partial\Psi_3}{\partial z}-\frac{\partial V(\phi,T)}{\partial \phi}
\label{ELmain}
\end{equation}

\begin{equation}
\frac{\partial \Psi_2}{\partial t}=\frac{\partial \Psi_1}{\partial r}
\end{equation}

\begin{equation}
\frac{\partial \Psi_3}{\partial t}=\frac{\partial \Psi_1}{\partial z}
\label{el4}
\end{equation}

 The implementation 
of the scheme is as follows. When we have a 1D non-linear hyperbolic equation of the form 

\begin{equation}
\frac{\partial u}{\partial t}+\frac{\partial f(u)}{\partial x} = 0
\end{equation}
the finite difference form given by the Lax Friedrichs scheme is
 
 \begin{equation}
 u_i^{n+1}=\frac{1}{2}(u_{i+1}^{n}+u_{i-1}^{n}) - \frac{\Delta t}{2 \Delta x}(f(u_{i+1}^{n})-f(u_{i-1}^{n}))
 \end{equation}
  The superscript $n$, $n+1$ denotes the time steps separated by $\Delta t$ and the 
 subscripts $i-1$, $i$ and $i+1$ denote points in space separated by $\Delta x$. This scheme
 is first order accurate and suffers from numerical dissipation problem. In our
 case, this scheme along with a small dissipation term in the field evolution 
 Eq.(\ref{ELmain}) at the initial times of the evolution seems to work well, as a firt step. 
If the effective potential rapidly changes due to the 
change in temperature profile, the field  could have large oscillations but we expect that in the presence of a medium which is coupled to the Polyakov loop field, this change will be smoother.
The dissipation term, in a way, takes care of this coupling.
 The dissipation term is space dependent with its magnitude 
proportional to the magnitude of the Polyakov loop field. Hence it is small at the edges and does 
not hinder the expansion. This term is also gradually 
decreased to zero after a few time steps since as the temperature profile expands, the field profile 
seems to stabilize and do not oscillate rapidly. The surface term starts to affect the velocity build 
up only later in the evolution and hence this also makes sure that the surface effects are not affected
 by the dissipation term. If indeed there are large oscillations in 
the field, this scheme along with KT cannot handle the numerical simulation.
Even with no oscillations, numerical instabilities develop at later times ($>$ 15 fm) at the boundary 
because of the source term and a better scheme for the scalar field evolution may improve this.

 To summarize the numerical implementation, we use an initial Gaussian energy density profile 
 to calculate the corresponding pressure, temperature and Polyakov loop field profiles. We 
assume zero initial velocity of the fluid and also $\frac{d\phi}{dt}=0$.  Using the initial 
temperature and $\phi$  profiles, Eq.(\ref{EulLag2}) is advanced by a time step. 
 The energy-momentum tensor of the field is calculated  and Eq.(\ref{consrv2}) is advanced by a time step using KT scheme and a second order RK method.  Energy density, pressure, velocity and temperature of the fluid are calculated from the new $T^{\mu\nu}_{fluid}$ and the new temperature profile is used to evolve Eq.(\ref{EulLag2}) again.

\section{Results and discussions}
\label{sec:results}

We study the development of flow with time in central and non-central initial conditions.
The radial flow velocity and momentum anisotropy are defined as in \cite{kolb}.
The average radial flow velocity is given by 

\begin{equation}
\left<v_r\right> =\frac{\left<\gamma\sqrt{v_x^2+v_y^2}\right>}{\left<\gamma\right>} 
\end{equation}
The angular brackets denote energy density weighted averages. The momentum anisotropy 
is given by

\begin{equation}
\epsilon_p = \frac{\left<T^{xx}-T^{yy}\right>}{\left<T^{xx}+T^{yy}\right>}
\end{equation}

\begin{figure}[h]
  \centering
  {\rotatebox{0}{\includegraphics[width=0.85\hsize]
      {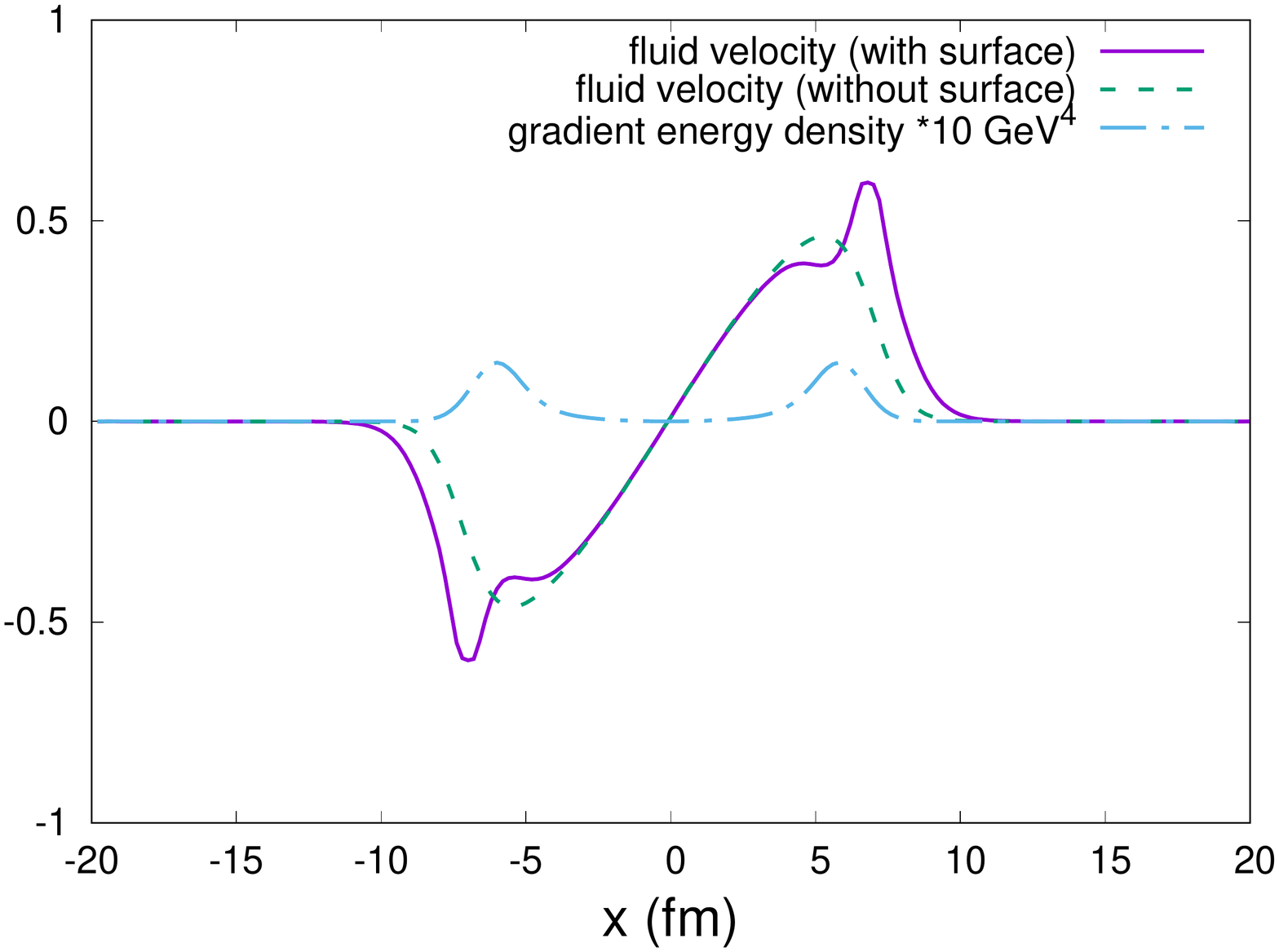}}
  }
\caption{Velocity profile along x-axis with (the continuous curve) and without (the dashed curve) surface tension at 2 fm/c. The dot-dashed curve represents the gradient energy contribution from the Polyakov
loop field and is magnified by a factor 10 here.}
    \label{vel}
\end{figure}

We see that the presence of the surface tension slows down the buildup of radial flow
velocity as well as momentum anisotropy. It can be easily understood using Eq.(\ref{consrv2}).
As the fluid expands the gradient energy of the field increases as a function of surface
area, so the fluid has to do work to compensate for this. If we plot the velocity profile 
of the fluid, we see that the fluid slows down near the surface thereby slowing
down the build up of radial flow velocity. Figure \ref{vel} shows the velocity profiles with
 and without surface tension. The dot-dashed curve represents the
surface tension (gradient energy of $\phi$ field). Since this value is very small, it is magnified 
by a factor of 10 here.
One can see here that the fluid velocity slows down near the surface. Even though it picks up
outside the surface, the energy density there is negligible to contribute to the flow.

\begin{figure}[h]
  \centering
  \subfigure[]
  {\rotatebox{0}{\includegraphics[width=0.4\hsize]
      {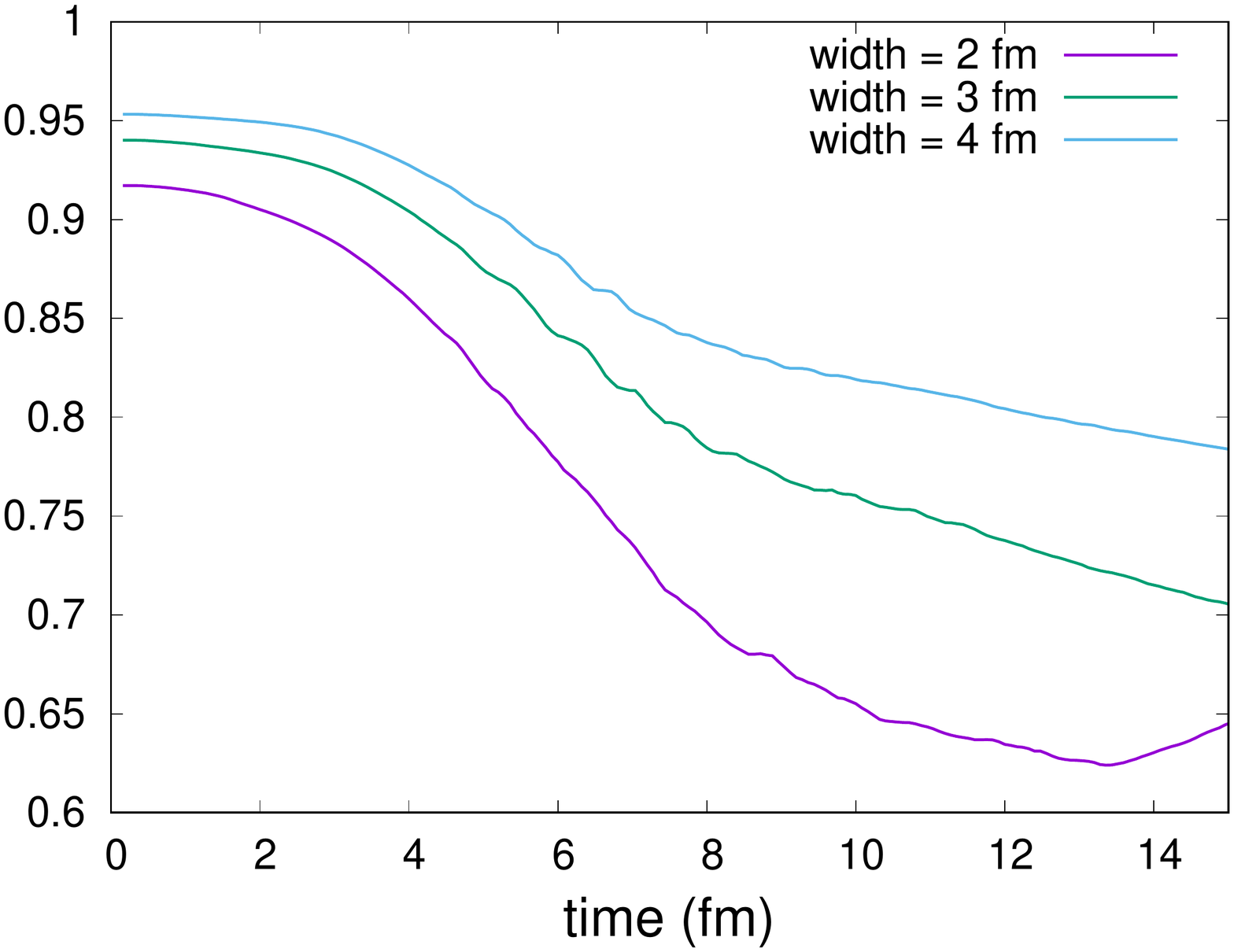}}
      \label{r2}
  }
  \subfigure[]
  {\rotatebox{0}{\includegraphics[width=0.4\hsize]
      {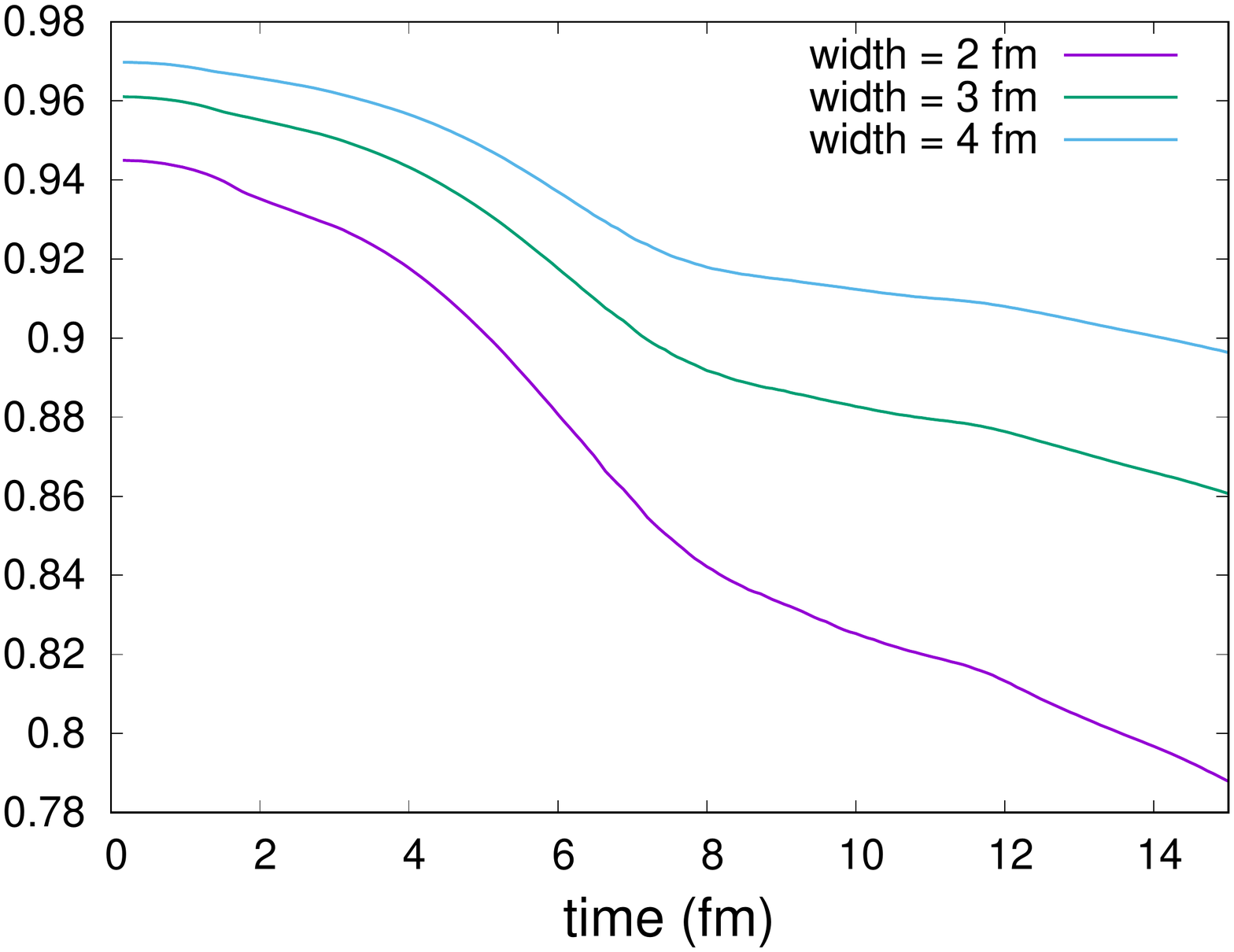}}
      \label{r3}
  }
  \subfigure[]
  {\rotatebox{0}{\includegraphics[width=0.4\hsize]
      {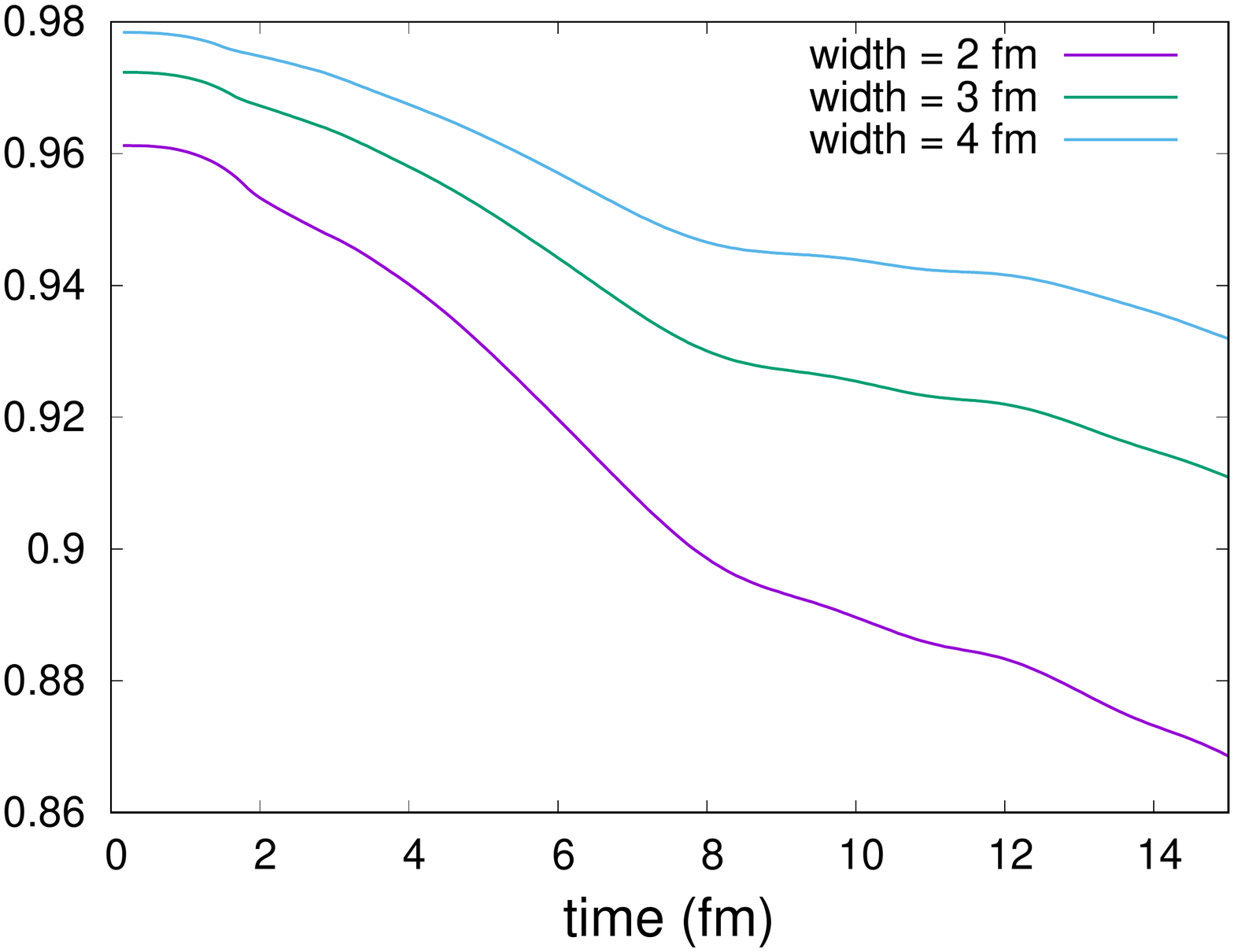}}
      \label{r4}
  }
  
\caption{Evolution of the ratio of radial flow with and without phase boundary effect. 
Figures (a), (b) and (c) show the evolution of radial 
flow for spherically symmetric Gaussian initial conditions with maximum energy 
values 2, 3 and 4 GeV/fm$^3$ respectively. The three curves in each case show the evolution for 
different sizes of the Gaussian. }
\end{figure}

\begin{figure}[h]
  \centering
  
  \subfigure[]
  {\rotatebox{0}{\includegraphics[width=0.4\hsize]
      {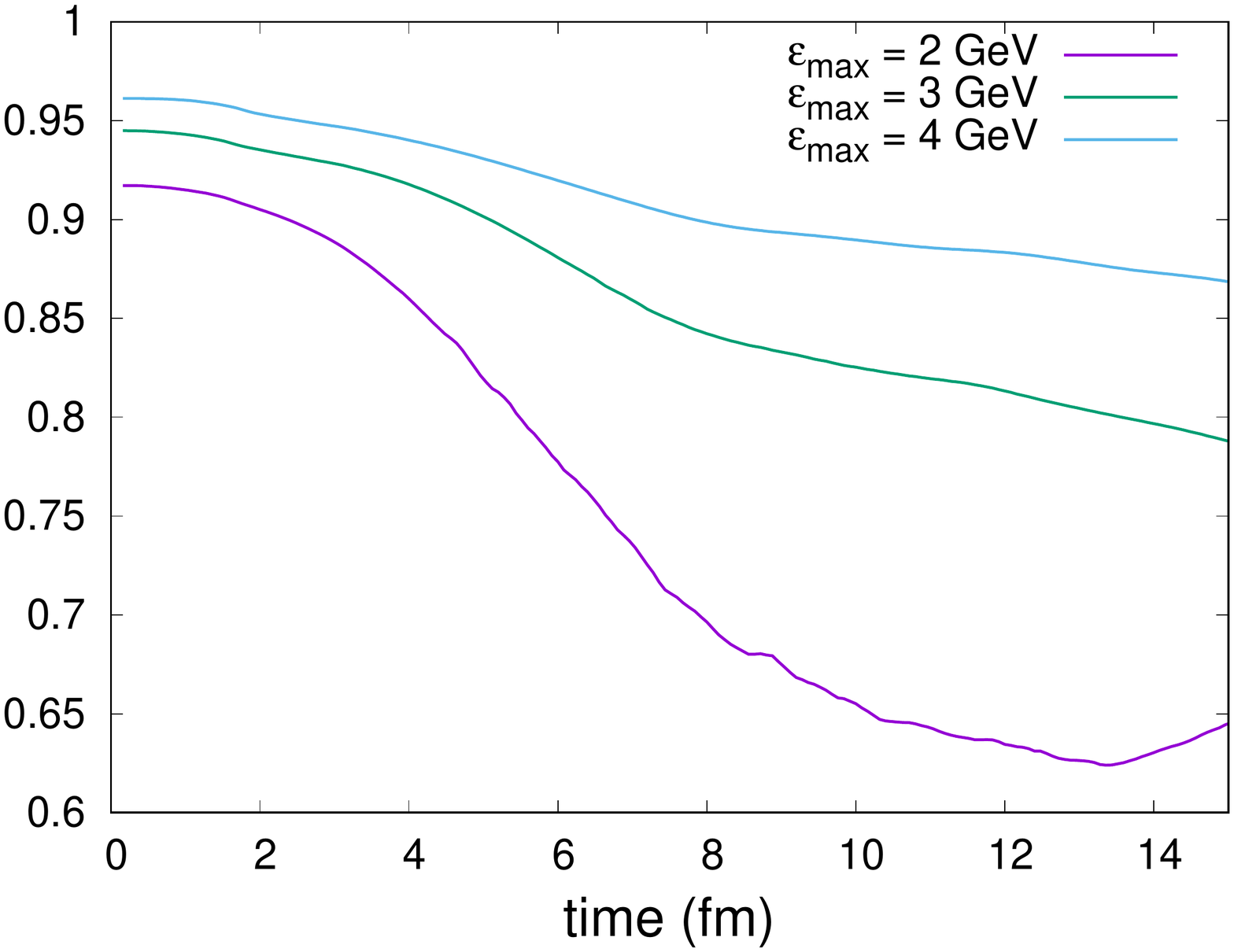}}
      \label{e2}
  }
  \subfigure[]
  {\rotatebox{0}{\includegraphics[width=0.4\hsize]
      {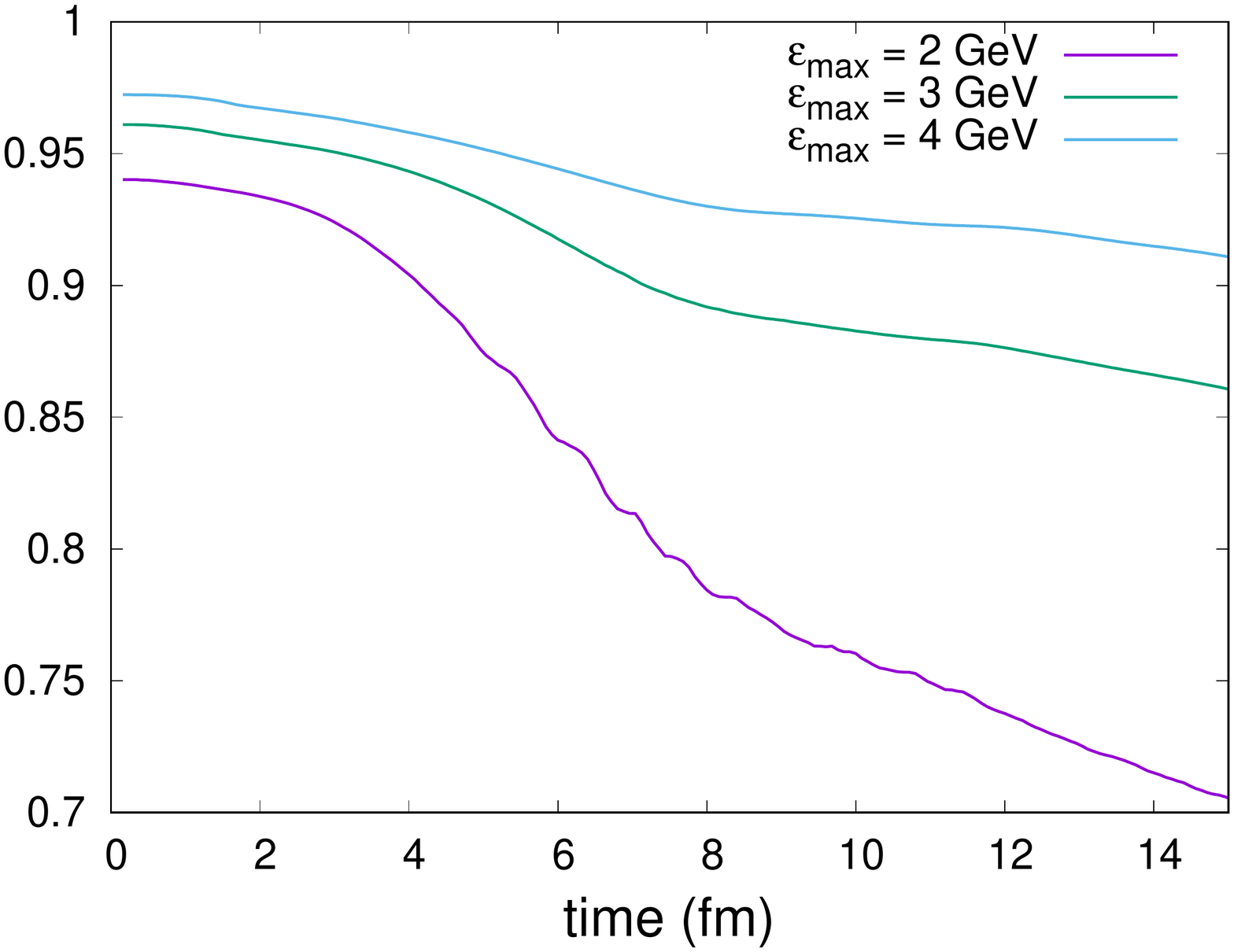}}
      \label{e3}
  }
  \subfigure[]
  {\rotatebox{0}{\includegraphics[width=0.4\hsize]
      {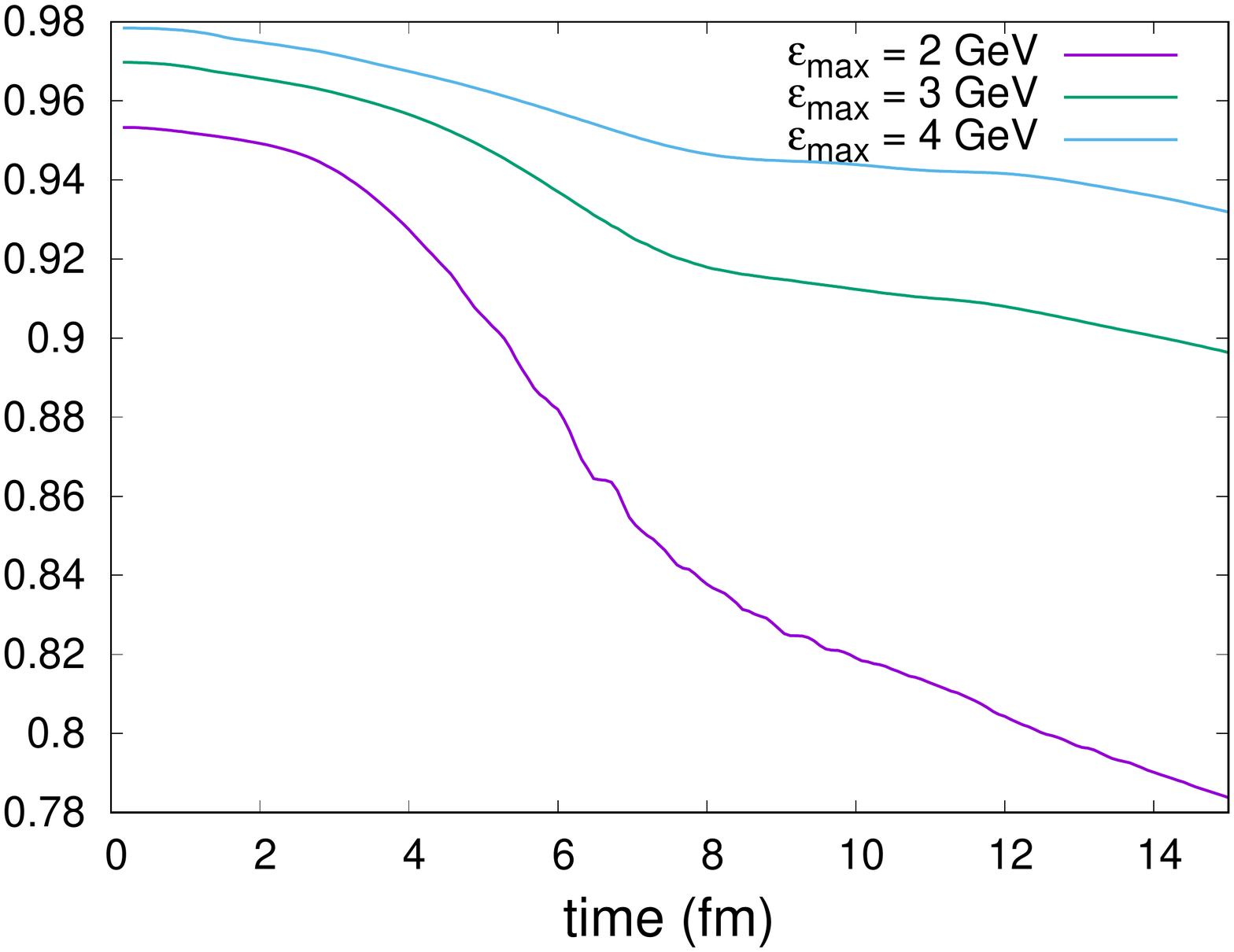}}
      \label{e4}
  }
\caption{Evolution of the ratio of radial flow with and without phase boundary effect. 
Figures (a), (b) and (c) show the evolutions
for 3 different sizes of the Gaussian - 2, 3 and 4 fm respectively with 3 curves in each case
showing the energy dependence as indicated in the figures.}
\end{figure}

Figures below show the evolution of the ratio of radial flow velocities with and without 
the order parameter field. We used 3 different widths for the spherically  symmetric 
Gaussian profiles and 3 different values for the maximum energy density.

Figures \ref{r2}, \ref{r3} and \ref{r4} show the evolution of radial flow at widths 2, 3 and 4 fm respectively,
the three curves in each corresponding to three values of maximum energy densities. As expected, the suppression of 
radial flow decreases with increase in size. This is explained by the fact that as size
increases the ratio of surface area to volume decreases and the flow build up or the velocity
build up which is a volume effect becomes stronger faster than the surface tension effect.
Figures \ref{e2}, \ref{e3} and \ref{e4} show the evolution of radial flow at maximum initial energy densities
2, 3, 4 GeV/fm$^3$ respectively. The three curves are the results for 3 different widths.
Again, as initial energy density increases, the surface tension effect decreases since with
increase in energy density, the pressure of the plasma increases faster than the surface energy 
and this strengthens the radial flow. What is of the most interest here is the percentage of 
suppression. At 15 fm/c, or an initial radius of 2 fm and an initial energy density 2 GeV/fm$^3$
gives a suppression as large as 40\%. It gradually decreases with increase in size and energy,
but there is a 5\% suppression even at 4~fm and 4~Gev/fm$^3$.

\begin{figure}[h]
  \centering
  \subfigure[]
  {\rotatebox{0}{\includegraphics[width=0.4\hsize]
      {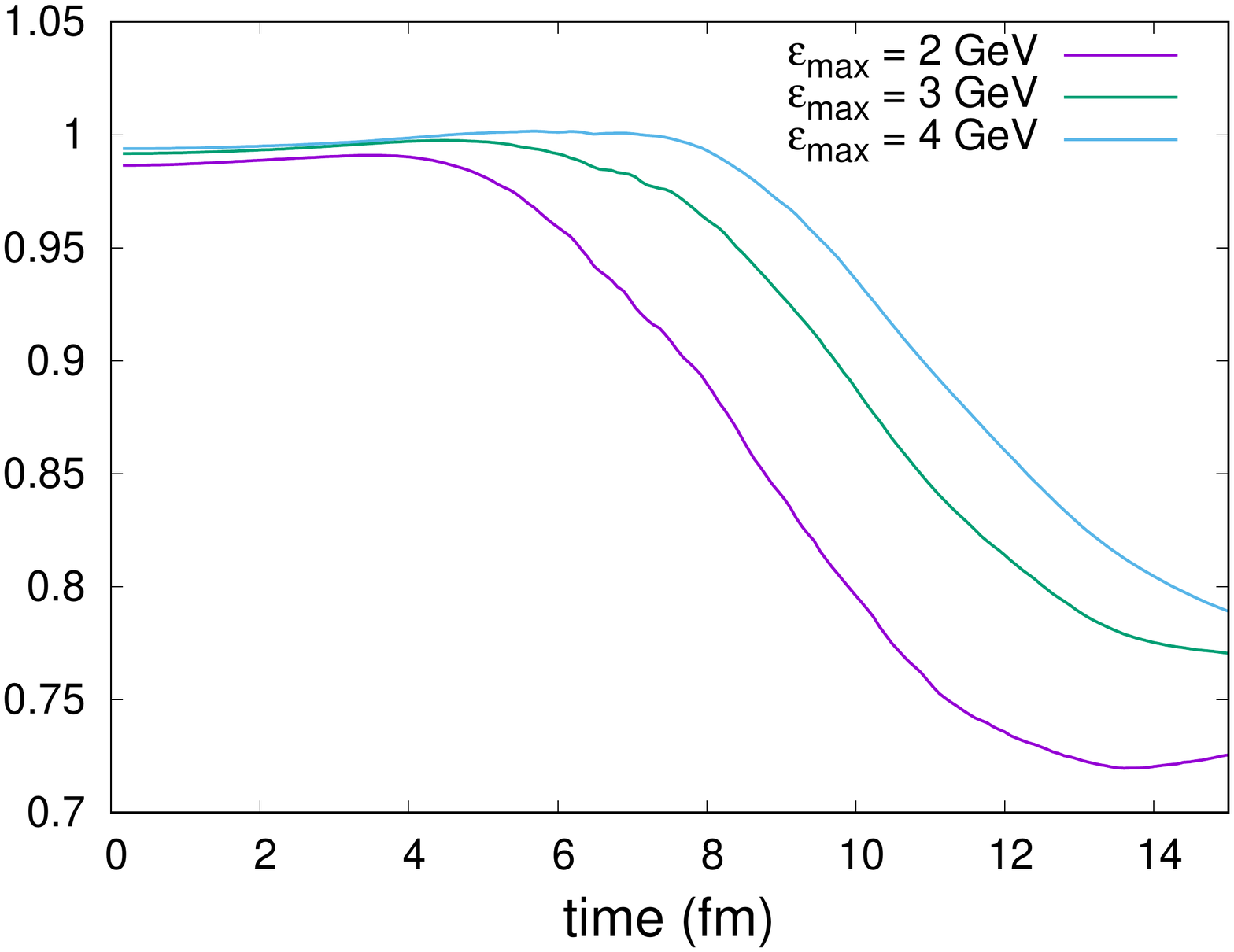}}
      \label{ep23}
  }
  \subfigure[]
  {\rotatebox{0}{\includegraphics[width=0.4\hsize]
      {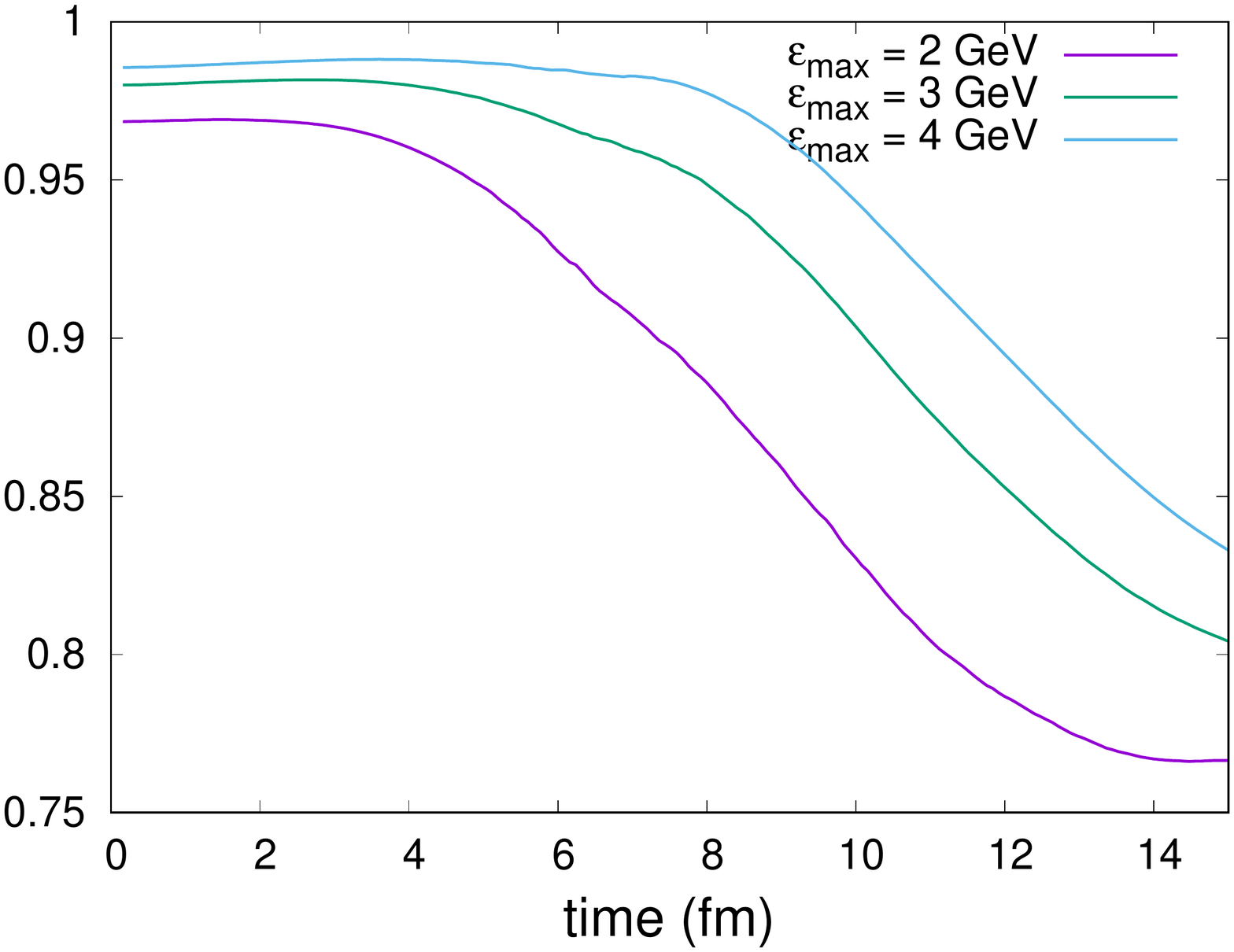}}
      \label{ep24}
  }
  \subfigure[]
  {\rotatebox{0}{\includegraphics[width=0.4\hsize]
      {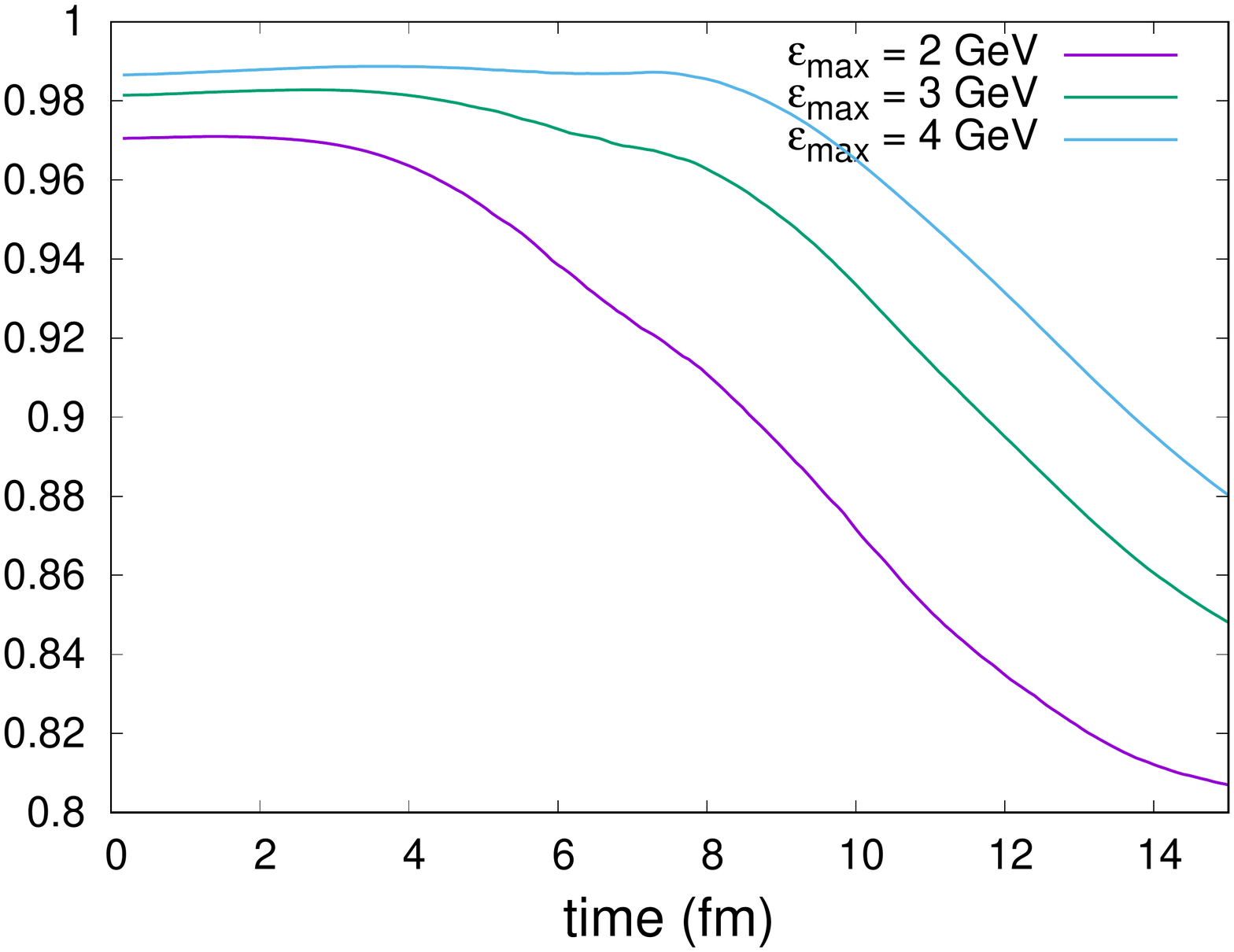}}
      \label{ep25}
  }

\caption{Evolution of ratio of momentum anisotropy with and without phase boundary. 
Figures (a), (b) and (c) have major axes
3 fm, 4 fm and 5 fm respectively for the ellipsoid when the minor axis is 2 fm for all.
The three different curves show the evolution for 3 different maximum energy values
as indicated in the figures for the initial Gaussian energy density profiles.}
\end{figure}

Similar results are seen in the development of momentum anisotropy as well.
Figures \ref{ep23}, \ref{ep24} and \ref{ep25} show the evolution of momentum 
anisotropy with the minor axis of the ellipsoid 2 fm and the major axis 
3, 4 and 5 fm respectively. It is interesting to note here that, when the major axis
increases from 3 fm to 5 fm, the suppression of the momentum anisotropy decreases from 25\%
to 20\% only, for $\epsilon_{max} = 2$ GeV. This is because even though an increase in major 
axis increases the size of the system, it also increases the anisotropy. This result also
indicates that the effect of the phase boundary will be reflected in the development
of elliptic flow.

\section{Summary and outlook}
\label{sec:summ}

We introduce a new way of studying the effect of the presence of a surface
in hydrodynamic evolution by coupling it to the order parameter field evolution. 
Polyakov loop model  with a weak first order transition
and no quarks is used for the equation of state of the fluid as well as the
equation of motion of the order parameter field, whose gradient energy represents
the phase boundary. 
The results show that even in the presence of a weak first order transition, the surface
tension at the phase boundary tends to slow down the expansion. This effect is
stronger when the system size is small or if the initial energy density of the system is small. 

Even in case of a cross over transition, there will be a non-zero gradient energy 
at the phase boundary. We wrote down an effective potential with a 
small explicit symmetry breaking and a cross over transition and checked 
the effects of the gradient term on hydrodynamics. The gradient term was small to make any visible 
effects in our case. But if the cross over is sharper, the flow may be affected.
We also tried to study the effect of a stronger first order transition by scaling
the effective potential to increase the barrier height between the two states
at transition temperature. A higher barrier height means higher surface energy and
as a result more resistance to expansion.
With our parametrisation, we found that the effect of this on flow
was smaller since the higher barrier height made the surface sharper and as a result 
the fluid started feeling its presence only later. But we expect a stronger suppression at
later times.
In order to study it more accurately,
one needs a more realistic model of strong first order transition as well as 
a more accurate numerical method for solving the order parameter equation of motion so that
the system can be evolved for longer times.

We have shown that the gradient energy in the order parameter field at a phase boundary affects 
collective flow.
This effect is more for smaller systems and at lower energies in presence
of a weak first order transition. The model studied here assumes a baryon less fluid. First order transition is
expected to occur at high baryon densities and it will be interesting to study 
this effect including the flow of baryons in a more realistic case. One should also 
address the presence of hadrons using a better equation of state. In low energy collisions,
the system is expected to spend some time in hadronic state before freeze out and it will be 
interesting to see the effect of this in the suppressed fluid flow. But we believe that our study
is an effective way to capture the dynamics of an expanding fluid with a phase boundary. 
The complex Polyakov loop field evolution in an effective potential can also
 explain the dynamics of center domains, Z(3) walls and strings~\cite{rkmams1,rkmams2,rkmamsfo} and we believe
that our formulation with appropriately modified order parameter field equations and 
equations of state could also study the dynamics of Z(3) walls and strings in a quark
gluon fluid - the effect of topological structures on flow and vice versa.

\begin{acknowledgments}
We are very grateful to Andrey Khvorostukhin for detailed discussions of the numerical techniques,
 suggesting KT scheme for the conservation equations as well as for his comments on the manuscript. We also thank Ajit Srivastava for very useful comments and suggestions on this work. We thank Shreyansh S. Dave for his useful comments on the manuscript.
 \end{acknowledgments}

\end{document}